\documentclass[preprintnumbers,showpacs,amsmath,amssymb,floatfix,12pt,superscriptaddress,nofootinbib]{revtex4}
\usepackage{graphicx}
\usepackage{epsfig}
\usepackage{bm}
\usepackage{amsfonts}

\begin{document}

\title{Variable $G$ correction to statefinder parameters of dark energy}

\author{Mubasher Jamil}
\email{mjamil@camp.nust.edu.pk} \affiliation{Center for Advanced
Mathematics and Physics, National University of Sciences and
Technology, H-12, Islamabad, Pakistan}

\begin{abstract}
\vspace*{1.5cm} \centerline{\bf Abstract} \vspace*{1cm} Motivated by
several observational and theoretical developments concerning the
variability of Newton's gravitational constant with time $G(t)$, we
calculate the varying $G$ correction to the statefinder parameters
for four models of dark energy namely interacting dark energy
holographic dark energy, new-agegraphic dark energy and generalized
Chaplygin gas.\\

\end{abstract}


 \maketitle

\newpage
\section{Introduction}

Recent cosmological observations obtained by SNe Ia {\cite{c1}},
WMAP {\cite{c2}}, SDSS {\cite{c3}} and X-ray {\cite{c4}} indicate
that the observable universe experiences an accelerated expansion.
Although the simplest way to explain this behavior is the
consideration of a cosmological constant \cite{c7}, the known
fine-tuning problem \cite{8} led to the dark energy paradigm. The
dynamical nature of dark energy, at least in an effective level, can
originate from a variable cosmological ``constant'' \cite{varcc}, or
from various fields, such is a canonical scalar field (quintessence)
\cite{quint}, a phantom field, that is a scalar field with a
negative sign of the kinetic term \cite{phant}, or the combination
of quintessence and phantom in a unified model named quintom
\cite{quintom}. Finally, an interesting attempt to probe the nature
of dark energy according to some basic quantum gravitational
principles is the holographic dark energy paradigm \cite{holoext}.

Dirac's Large Numbers Hypothesis (LNH) is the origin of many
theoretical studies of time-varying $G$. According to LNH, the value
of $\dot{G} /G$ should be approximately the Hubble rate \cite{lnh}.
Although it has become clear in recent decades that the Hubble rate
is too high to be compatible with experiments, the enduring legacy
of Dirac's bold stroke is the acceptance by modern theories of
non-zero values of $\dot{G} /G$ as being potentially consistent with
physical reality. Moreover, it is the task of a final quantum
gravity theory to answer how, why and what a particular physical
constant (like $G$) takes a specific value. As we are considering
only variations of $G$ with respect to temporal coordinate $t$,
there is a suggestion to take $G(r)$, as a spatial (or radial)
dependent quantity \cite{milk}. In \cite{sri}, it is shown that $G$
was actually varying before the electroweak era in the early
Universe, while after the spontaneous symmetry breaking, the
gravitational coupling $G$ attained a finite constant value known
today. This line of reasoning can be extended to other types of
couplings in particle physics as well \cite{dent} but we shall not
go into that. In the same manner, some authors \cite{debnath} have
also considered the speed of light as a time varying quantity.

There have been many proposals in the literature attempting to
theoretically justified a varying gravitational constant, despite
the lack of a full, underlying quantum gravity theory. Starting with
the simple but pioneering work of Dirac \cite{Dirac:1938mt}, the
varying behavior in Kaluza-Klein theory was associated with a scalar
field appearing in the metric component corresponding to the  $5$-th
dimension \cite{kal} and its size variation \cite{akk}. An
alternative approach arises from Brans-Dicke framework \cite{bd},
where the gravitational constant is replaced by a scalar field
coupling to gravity through a new parameter, and it has been
generalized to various forms of scalar-tensor theories \cite{gen},
leading to a considerably broader range of variable-$G$ theories. In
addition, justification of a varying Newton's constant has been
established with the use of conformal invariance and its induced
local transformations \cite{bek}. Finally, a varying $G$ can arise
perturbatively through a semiclassical treatment of Hilbert-Einstein
action \cite{19}, non-perturbatively through quantum-gravitational
approaches within the ``Hilbert-Einstein truncation'' \cite{21}, or
through gravitational holography \cite{Guberina,7}.

There is convincing evidence of a varying Newton's constant $G$:
Observations of Hulse-Taylor binary pulsar B$1913+16$ gives a
following estimate $0<\dot{G}/G\sim2\pm4\times10^{-12}{yr}^{-1}$
\cite{kogan}, helioseismological data gives the bound
$0<\dot{G}/G\sim1.6\times10^{-12}{yr}^{-1}$ \cite{guenther} (see Ref
\cite{ray1} for various bounds on $\dot{G}/G$ from observational
data). The variability in $G$ results in the emission of
gravitational waves. In another approach, it is shown that $G$ can
be oscillatory with time \cite{pradhan}. It is recently proposed
that variable cosmic constants are coupled to each other i.e.
variation in one leads to changes in others \cite{vishwakarma}. A
variable gravitational constant also explains the dark matter
problem as well \cite{goldman}. Also discrepancies in the value of
Hubble parameter can be removed with the consideration of variable
$G$ \cite{bertolami1}. Motivated by the above arguments, we shall
calculate the the varying $G$ corrections to the statefinder
parameters for four chosen models of dark energy in the coming
section and subsections.

\section{The model}
The action of our model is \cite{kho}
\begin{equation}\label{1a}
S=\int d^4x\mathfrak{L}=\int d^4x\Big\{ \sqrt{-g}\Big[
\frac{R}{G}+F(G) \Big]+\mathfrak{L_m} \Big\},
\end{equation}
where $G$ is the Newton's gravitational constant and $F(G)$ is an
arbitrary function of $G$ while $\mathfrak{L_m}$ is a matter
Lagrangian. From the Euler-Lagrange equation
\begin{equation}\label{1b}
\frac{\partial \mathfrak{L}}{\partial G}=\nabla_\mu\frac{\partial
\mathfrak{L}}{\partial(\partial_\mu G)},
\end{equation}
to obtain
\begin{equation}\label{1c}
\frac{\partial F}{\partial G}=\frac{R}{G^2}.
\end{equation}
Moreover from the variation with respect to $g_{\mu\nu}$, we have
from (\ref{1a}),
\begin{equation}\label{1d}
R_{\mu\nu}-\frac{1}{2}g_{\mu\nu}R=8\pi G
T_{\mu\nu}+g_{\mu\nu}\Big(\frac{1}{2}GF(G)\Big).
\end{equation}
Writing
\begin{equation}\label{1e}
\frac{1}{2}GF(G)=\Lambda(t),
\end{equation}
we get from (\ref{1d})
\begin{equation}\label{1f}
R_{\mu\nu}-\frac{1}{2}g_{\mu\nu}R=8\pi G(t)
T_{\mu\nu}+g_{\mu\nu}\Lambda(t).
\end{equation}

Sahni et al \cite{sahni} introduced a pair of cosmological
diagnostic pair $\{r,s\}$ which they termed as Statefinder. The two
parameters are dimensionless and are geometrical since they are
derived from the cosmic scale factor alone, though one can rewrite
them in terms of the parameters of dark energy and matter.
Additionally, the pair gives information about dark energy in a
model independent way i.e. it categorizes dark energy in the context
of background geometry only which is not dependent on the theory of
gravity. Hence geometrical variables are universal. Also this pair
generalizes the well-known geometrical parameters like the Hubble
parameter and the deceleration parameter. This pair is algebraically
related to the equation of state of dark energy and its first time
derivative.

The statefinder parameters were introduced to characterize primarily
flat universe ($k=0$) models with cold dark matter (dust) and dark
energy. They were defined as
\begin{equation}\label{1}
r\equiv\frac{\dddot a}{aH^3},
\end{equation}
\begin{equation}\label{2}
s\equiv \frac{r-1}{3(q-\frac{1}{2})}.
\end{equation}
Here $q=-\frac{\ddot a}{aH^2}$ is the deceleration parameter.

For cosmological constant with a fixed equation of state ($w=-1$)
and a fixed Newton's gravitational constant, we have $\{1,0\}$.
Moreover $\{1,1\}$ represents the standard cold dark matter model
containing no radiation while Einstein static universe corresponds
to $\{\infty,-\infty\}$ \cite{debnath23}. In literature, the
diagnostic pair is analyzed for various dark energy candidates
including holographic dark energy \cite{zhang}, agegraphic dark
energy \cite{wei}, quintessence \cite{zhang1}, dilaton dark energy
\cite{dilaton}, Yang-Mills dark energy \cite{yang}, viscous dark
energy \cite{vis}, interacting dark energy \cite{pavon}, tachyon
\cite{shao}, modified Chaplygin gas \cite{debnath1} and $f(R)$
gravity \cite{song} to name a few.

For the present homogeneous, isotropic and spatially flat universe
containing dark energy and dark matter (ignoring baryonic matter,
radiation, neutrinos etc), the Friedmann equation is
\begin{equation}\label{3}
H^2=\frac{8\pi G}{3}(\rho_m+\rho_x)
\end{equation}
Let us first consider dark energy obeying an equation of state of
the form $p_x=w\rho_x$. The formalism of Sahni and coworkers
\cite{sahni} will be generalized to permit varying gravitational
constant. In this case, the definition of $s$ is generalized to
\begin{equation}\label{4}
s\equiv \frac{r-\Omega}{3(q-\frac{\Omega}{2})}.
\end{equation}
Here $\Omega=\Omega_m+\Omega_x$. The deceleration parameter may be
expressed as
\begin{equation}\label{5}
q=\frac{1}{2}[\Omega_m+(1+3w)\Omega_x].
\end{equation}
Hence if $\Omega_x$, $\Omega$ and $q$ are determined by measurements
the equation of state factor $w$ may be found from
\begin{equation}\label{6}
w=\frac{2q-\Omega}{3\Omega_x}.
\end{equation}
Differentiation of parameter $q$, together with (\ref{1}) leads to
\begin{equation}\label{7}
r=2q^2+q-\frac{\dot q}{H}.
\end{equation}
From (\ref{5}), we have
\begin{equation}\label{8}
\dot q=\frac{1}{2}\dot{\Omega}_m+\frac{1}{2}(1+3w)\dot{\Omega}_x+
\frac{3}{2}\dot w\Omega_x.
\end{equation}
Furthermore,
\begin{equation}\label{9}
\dot{\Omega}=\frac{\dot\rho}{\rho_{\text{cr}}}
-\frac{\rho}{\rho_{\text{cr}}^2}\dot{\rho}_{\text{cr}},
\end{equation}
with
\begin{equation}\label{10}
\dot{\rho}_{\text{cr}}=\rho_{\text{cr}}\Big( 2\frac{\dot
H}{H}-\frac{\dot G}{G} \Big),
\end{equation}
and
\begin{equation}\label{11}
\dot H=-H^2(1+q).
\end{equation}
Hence
\begin{equation}\label{12}
\dot{\rho}_{\text{cr}}=-H\rho_{\text{cr}}[2(1+q)+\Delta_G],
\end{equation}
where $\Delta_G\equiv G'/G$, $\dot G =HG'$ which leads to
\begin{equation}\label{13}
\dot{\Omega}=\frac{\dot\rho}{\rho_{\text{cr}}}+ \Omega
H[2(1+q)+\Delta_G].
\end{equation}
For cold dark matter $\dot{\rho}_m=-3H\rho_m$, (\ref{13}) gives
\begin{equation}\label{14}
\dot{\Omega}_m=\Omega_m H(-1+2q+\Delta_G),
\end{equation}
and for dark energy $\dot{\rho}_x=-3(1+w)H\rho_x$, (\ref{13}) gives
\begin{equation}\label{15}
\dot{\Omega}_x=\Omega_x H(-1-3w+2q+\Delta_G).
\end{equation}
Inserting (\ref{14}) and (\ref{15}) into (\ref{8}) and the resulting
expression into (\ref{7}) finally leads to
\begin{equation}\label{16}
r=\Omega_m+\Big[ 1+\frac{9}{2}w(1+w) \Big]\Omega_x-\frac{3}{2H}\dot
w\Omega_x-\frac{\Delta_G}{2}[\Omega_m+(1+3w)\Omega_x].
\end{equation}
Inserting the expression (\ref{16}) into (\ref{4}) leads to
\begin{equation}\label{17}
s=1+w-\frac{\dot
w}{3wH}-\frac{\Delta_G}{9w\Omega_x}[\Omega_m+(1+3w)\Omega_x]
\end{equation}
Assuming a spatially flat universe satisfying $\Omega_m=1-\Omega_x,$
we obtain from (\ref{16}) and (\ref{17}), the following expressions
\begin{equation}\label{18}
r= 1+\frac{9}{2}w(1+w)\Omega_x-\frac{3}{2H}\dot
w\Omega_x-\frac{\Delta_G}{2}(1+3w\Omega_x),
\end{equation}
\begin{equation}\label{19}
s=1+w-\frac{\dot w}{3wH}-\frac{\Delta_G}{9w\Omega_x}(1+3w\Omega_x).
\end{equation}
Note that (\ref{18}) and (\ref{19}) provide necessary variable $G$
corrections to statefinder parameters of Sahni et al \cite{sahni}.
In particular for the $\Lambda$CDM model, the above pair gives
\begin{equation}\label{s1}
r= -\frac{\Delta_G}{2}(1-3\Omega_x),
\end{equation}
\begin{equation}\label{s2}
s=\frac{\Delta_G}{9\Omega_x}(1-3\Omega_x).
\end{equation}
Moreover the standard cold dark matter corresponds to
$\{-\Delta_G/2,-\infty\}$. For a universe containing only dark
energy i.e. $\Omega_m=0$, we get from (\ref{16}) and (\ref{17}):
\begin{equation}\label{20}
r=\Big[ 1+\frac{9}{2}w(1+w) \Big]\Omega_x-\frac{3}{2H}\dot
w\Omega_x-\frac{\Delta_G}{2}(1+3w)\Omega_x,
\end{equation}
\begin{equation}\label{21}
s=1+w-\frac{\dot w}{3wH}-\frac{\Delta_G}{9w}(1+3w).
\end{equation}

\subsection{Interacting dark energy}

There is a class of cosmological models in which evolution of the
universe depends on the interaction of cosmic components like dark
energy and dark matter (and possibly radiation and neutrinos).
Models in which the main energy components do not evolve separately
but interact with each other bear a special interest since they may
alleviate or even solve the ``cosmic coincidence problem''. The
problem can be summarily stated as ``why now?'', that is to say:
``Why the energy densities of the two main components happen to be
of the same order today?''

The conservation equations for the interacting dark energy-dark
matter are
\begin{eqnarray}\label{42}
\dot {\rho}_x+3H\rho_x(1+w)&=&-Q,\nonumber\\
\dot {\rho}_m+3H\rho_m&=&Q.
\end{eqnarray}
The coupling between dark components could be a major issue to be
confronted in studying the physics of dark energy. However, so long
as the nature of these two components remain unknown it will not be
possible to derive the precise form of the interaction from first
principles. Therefore, one has to assume a specific coupling from
the outset \cite{Ads,Ame2} or determine it from phenomenological
requirements \cite{pavon}. Here we take $Q=-3\Pi H$ which measures
the strength of the interaction \cite{pavon}. For later convenience
we will write it as $Q=-3\Pi H$ where the new quantity $\Pi$ has the
dimension of a pressure.

The second Friedmann equation is
\begin{equation}\label{43}
\dot H=-4\pi G(\rho_t+p),\ \
\end{equation}
where $\rho_t=\rho_m+\rho_x$. Notice that if the cosmic energy
density and pressure violate the null energy condition $\rho_t+p<0$,
then $\dot H>0$, i.e. the Hubble horizon $H^{-1}$ will be shrinking.
This situation can arise in a phantom-type dark energy model. Also
if $\rho_t+p>0$ then the Hubble horizon is expanding and this can be
due to a quintessence-like dark energy. Moreover for a cosmological
constant dominated universe ($\rho_t+p=0$) or during inflation in
the early universe, the Hubble horizon is a fixed quantity.
Differentiating (\ref{43}) w.r.t $t$ and using (\ref{42}) and
(\ref{3}), we get
\begin{equation}\label{44}
\frac{\ddot
H}{H^3}=\frac{9}{2}\Big(1+\frac{p}{\rho_t}\Big)+\frac{9}{2}\Big[w(1+w)\frac{\rho_x}{\rho_t}
-\frac{w\Pi}{\rho_t}-\frac{\dot w
\rho_x}{3H\rho_t}\Big]-\frac{3}{2}\Delta_G\Big(1+\frac{p}{\rho_t}\Big).
\end{equation}
At variance with $H$ and $\dot H$, the second derivative $\ddot H$
does depend on the interaction between components. Consequently, to
discriminate between models with different interactions or between
interacting and non-interacting models, it is desirable to
characterize the cosmological dynamics additionally by parameters
that depend on $\ddot H$.

Consequently $r$ can be written in alternative form as
\begin{equation}\label{45}
r=\frac{\ddot H}{H^3}-3q+2.
\end{equation}
Using (\ref{44}) in (\ref{45}) and the expression of
$q=-\frac{1}{2}-\frac{3}{2}w\Omega_x(=-\frac{1}{2}-\frac{3p}{2\rho_t})$,
we obtain
\begin{equation}\label{46}
r=1+\frac{9}{2}\frac{w}{1+\kappa}\Big(
1+w-\frac{\Pi}{\rho_x}-\frac{\dot w}{3Hw} \Big)-\frac{3}{2}\Delta_G
\Big(1+\frac{w}{1+\kappa}\Big),
\end{equation}
where $\kappa\equiv\rho_m/\rho_x$ is a dimensionless but a varying
quantity. Note that a constant $\kappa$ require extreme fine-tuning.
Moreover for a fixed $\kappa$, the cosmic coincidence problem is
trivially resolved since the universe will remain in that state
forever. To resolve the coincidence problem, the density ratio
$\kappa$ has to be slowly varying over a time of the order $H^{-1}$.

Making use of (\ref{46}) in (\ref{2}), we obtain
\begin{equation}\label{47}
s= 1+w-\frac{\Pi}{\rho_x}-\frac{\dot
w}{3Hw}-\frac{\Delta_G}{3w}\Big(1+\kappa+w\Big),
\end{equation}
Note that for $\Delta_G=0$, the above expressions (\ref{46}) and
(\ref{47}) reduce to the expressions studied in \cite{pavon},
therefore (\ref{46}) and (\ref{47}) provide necessary variable $G$
corrections to the statefinder parameters for an interacting dark
energy model.

In a paper \cite{zim}, the authors showed that scaling solutions of
the form $\kappa\sim a^{-\xi}$, where $\xi$ denotes a constant
parameter in the range $[0,3]$ can be obtained when the dark energy
component decays into the pressureless matter. These solutions are
interesting because they alleviate the coincidence problem
\cite{jamil}. Indeed a model with $\xi=3$ amounts to the
$\Lambda$CDM model with $w=-1$ and $\Pi=0$. As mentioned earlier,
the $\xi=0$ trivially resolves the coincidence problem, which is of
no interest. Any solution $\xi<3$ renders the coincidence problem
less acute. In that scheme, with $w=$constant, it has been shown in
\cite{zim} that the interactions which produce scaling solutions are
given by
\begin{equation}\label{48}
\frac{\Pi}{\rho_x}=\Big( w+\frac{\xi}{3}
\Big)\frac{\kappa_0(1+z)^\xi}{1+\kappa_0(1+z)^\xi},
\end{equation}
where $\kappa=\kappa_0(1+z)^\xi$, $z=(a_0/a)-1$ is the redshift and
$\kappa_0\equiv\rho_{m0}/\rho_{x0}$ is the present density ratio.
Putting (\ref{48}) in (\ref{46}) and (\ref{47}), we get
\begin{equation}\label{49}
r=1+\frac{9}{2}\frac{w}{1+\kappa_0(1+z)^\xi}\Big[ 1+w-\Big(
w+\frac{\xi}{3}
\Big)\frac{\kappa_0(1+z)^\xi}{1+\kappa_0(1+z)^\xi}\Big]-\frac{3}{2}\Delta_G
\Big[1+\frac{w}{1+\kappa_0(1+z)^\xi}\Big],
\end{equation}
\begin{equation}\label{50}
s= 1+w-\Big( w+\frac{\xi}{3}
\Big)\frac{\kappa_0(1+z)^\xi}{1+\kappa_0(1+z)^\xi}-\frac{\Delta_G}{3w}\Big[1+w+\kappa_0(1+z)^\xi\Big].
\end{equation}

\subsection{Holographic dark energy}

The holographic dark energy is constructed in the light of the
holographic principle. Its framework is the black hole
thermodynamics \cite{BH22} and the connection (known from AdS/CFT
correspondence) of the UV cut-of of a quantum field theory, which
gives rise to the vacuum energy, with the largest distance of the
theory \cite{Cohen:1998zx}. Thus, determining an appropriate
quantity $L$ to serve as an IR cut-off, imposing the constraint that
the total vacuum energy in the corresponding maximum volume must not
be greater than the mass of a black hole of the same size, and
saturating the inequality, one identifies the acquired vacuum energy
as holographic dark energy:
\begin{equation}\label{22}
\rho_x=\frac{3c^2}{8\pi G}\frac{1}{L^2}.
\end{equation}
Here $c$ is the holographic parameter of order unity. Note that the
IR cut-off $L$ can be of several types such as particle horizon,
Hubble horizon and future event horizon. The definition of
holographic dark energy is sensitive to the choice of each horizon
and gives different dynamical features corresponding to each
horizon. For the present study, we don't need to chose a specific
length scale and work with general $L$.

The time evolution of (\ref{22}) is
\begin{equation}\label{23}
\dot{\rho}_x=-\rho_xH\Big( 2-\frac{2}{c}\sqrt{\Omega_x}+\Delta_G
\Big).
\end{equation}
Using (\ref{23}) in the energy conservation equation yields
\begin{equation}\label{24}
w=\frac{1}{3}\Big( -1-\frac{2}{c}\sqrt{\Omega_x}+\Delta_G \Big).
\end{equation}
Differentiating (\ref{24}), we obtain
\begin{equation}\label{25}
\frac{1}{H}\dot w\equiv\frac{d w}{d \ln
a}=-\frac{1}{3c}\sqrt{\Omega_x}(1-\Omega_x)\Big(
1+\frac{2}{c}\sqrt{\Omega_x}-\Delta_G \Big)+\frac{1}{3}\Delta'_G,
\end{equation}
where we have used
\begin{equation}\label{26}
\dot\Omega_x=H\Omega_x(1-\Omega_x)\Big(
1+\frac{2}{c}\sqrt{\Omega_x}-\Delta_G \Big).
\end{equation}
Putting (\ref{25}) and (\ref{26}) in (\ref{18}) and (\ref{19}), we
obtain
\begin{eqnarray}\label{27}
r&=&1+\Omega_x\Big( -1-\frac{2}{c}\sqrt{\Omega_x}+\Delta_G
\Big)\Big( 1-\frac{2}{c}\sqrt{\Omega_x}+\frac{3}{2}\Delta_G
\Big)+\frac{1}{2c}\Omega_x^{3/2}(1-\Omega_x)\Big(
1+\frac{2}{c}\sqrt{\Omega_x}-\Delta_G
\Big)\nonumber\\&&+\frac{1}{3}\Delta'_G\Omega_x-\frac{\Delta_G}{2}\Big[1+\Omega_x\Big(-
1-\frac{2}{c}\sqrt{\Omega_x}+\Delta_G \Big) \Big].
\end{eqnarray}
\begin{eqnarray}\label{28}
s&=&\frac{2}{3}-\frac{\sqrt{\Omega_x}}{c}-\frac{\Omega_x^{3/2}}{3c}+\frac{\Delta_G}{3}
+\frac{\Delta'_G}{3}\Big( 1+\frac{2}{c}\sqrt{\Omega_x}-\Delta_G
\Big)^{-1}+\frac{\Delta_G}{3\Omega_x}\Big(
1+\frac{2}{c}\sqrt{\Omega_x}-\Delta_G
\Big)^{-1}\nonumber\\&&\times\Big[1+\Omega_x\Big(-
1-\frac{2}{c}\sqrt{\Omega_x}+\Delta_G \Big) \Big].
\end{eqnarray}

\subsection{New-agegraphic dark energy}

An interesting attempt for probing the nature of dark energy (DE) is
the so-called ``agegraphic DE'' (ADE). This model was recently
proposed \cite{Cai1} to explain the acceleration of the universe
expansion within the framework of a fundamental theory such as
quantum gravity. The ADE model assumes that the observed DE comes
from the spacetime and matter field fluctuations in the universe.
Following the line of quantum fluctuations of spacetime, Karolyhazy
et al. \cite{Kar1} discussed that the distance $t$ in Minkowski
spacetime cannot be known to a better accuracy than $\delta{t}=\beta
t_{p}^{2/3}t^{1/3}$ where $\beta$ is a dimensionless constant of
order unity. Based on Karolyhazy relation and  Maziashvili arguments
\cite{Maz}, Cai proposed the original ADE model to explain the
acceleration of the universe expansion \cite{Cai1}. Since the
original ADE model suffers from the difficulty to describe the
matter-dominated epoch, a new model of ADE was proposed by Wei and
Cai \cite{Wei2}, while the time scale was chosen to be the conformal
time $\eta$ instead  of the age of the universe.

The definition of NADE is given by \cite{Wei2}
\begin{equation}\label{29}
\rho_x=\frac{3n^2}{8\pi G}\frac{1}{\eta^2},
\end{equation}
where $n$ is a constant of order unity. Its time evolution is
\begin{equation}\label{30}
\dot{\rho}_x=-H\rho_x\Big( \frac{2}{3na}\sqrt{\Omega_x}+\Delta_G
\Big).
\end{equation}
Making use of (\ref{30}) in the energy conservation equation yields
\begin{equation}\label{31}
w=-1+\frac{2}{3na}\sqrt{\Omega_x}+\frac{\Delta_G}{3}.
\end{equation}
Differentiating (\ref{31}) w.r.t. $t$ gives
\begin{equation}\label{32}
\frac{1}{H}\dot w=
\frac{1}{3na}\sqrt{\Omega_x}(1-\Omega_x)\Big(3-\frac{2}{3na}\sqrt{\Omega_x}-
\frac{\Delta_G}{3}\Big)-\frac{2}{3na}\sqrt{\Omega_x}+\frac{1}{3}\Delta'_G,
\end{equation}
where we have used
\begin{equation}\label{33}
\dot{\Omega}_x=H\Omega_x(1-\Omega_x)\Big(
3-\frac{2}{na}\sqrt{\Omega_x}-\Delta_G \Big).
\end{equation}
Using (\ref{31}) and (\ref{32}) in (\ref{18}) and (\ref{19}), we get
\begin{eqnarray}\label{34}
r&=&1+\frac{9}{2}\Omega_x\Big(
\frac{2}{3na}\sqrt{\Omega_x}+\frac{\Delta_G}{3} \Big)\Big(
-1+\frac{2}{3na}\sqrt{\Omega_x}+\frac{\Delta_G}{3} \Big)-
\frac{1}{2na}\Omega_x^{3/2}(1-\Omega_x)\Big(3-\frac{2}{3na}\sqrt{\Omega_x}-
\frac{\Delta_G}{3}\Big)\nonumber\\&&+\frac{1}{na}\Omega_x^{3/2}-\frac{1}{2}\Omega_x\Delta'_G
-\frac{\Delta_G}{2}\Big[1+\Omega_x(\Delta_G-3)+\frac{2}{na}\Omega_x^{3/2}\Big].
\end{eqnarray}
\begin{eqnarray}\label{35}
s&=&\frac{2}{3na}\sqrt{\Omega_x}+\frac{\Delta_G}{3}-\frac{1}{3}\Big(
-1+\frac{2}{3na}\sqrt{\Omega_x}+\frac{\Delta_G}{3}
\Big)^{-1}\Big[\frac{1}{3na}\sqrt{\Omega_x}(1-\Omega_x)\Big(3-\frac{2}{3na}\sqrt{\Omega_x}-
\frac{\Delta_G}{3}\Big)\nonumber\\&&-\frac{2}{3na}\sqrt{\Omega_x}+\frac{1}{3}\Delta'_G
\Big]-\frac{\Delta_G}{3\Omega_x(\Delta_G-3)+\frac{6}{na}\Omega_x^{3/2}}\Big[
1+ \Omega_x(\Delta_G-3)+\frac{2}{na}\Omega_x^{3/2} \Big].
\end{eqnarray}

\subsection{Generalized Chaplygin gas}

Interest in generalized Chaplygin gas (GCG) arose when it appeared
that it gives a unified picture of dark energy and dark matter. It
solely provides the density evolution of matter at high redshifts
and dark energy at low redshifts \cite{bilic}. Other successes of
GCG is that it explains the recent phantom divide crossing
\cite{zhang23}, is consistent with the data of type Ia supernova
\cite{senn} and the cosmic microwave background \cite{liu}. The GCG
emerges as an effective fluid associated with $d$-branes
\cite{borde} and can also be obtained from the Born-Infeld action
\cite{bento}. In the present context, we are treating GCG as a dark
energy candidate.

The equation of state for generalized Chaplygin gas is
\begin{equation}\label{36}
p_x=-\frac{A}{\rho_x^\alpha},
\end{equation}
where $\alpha$ and $A$ are constants. Using (\ref{36}) in the energy
conservation equation yields
\begin{equation}\label{37}
\rho_x=\Big( A+\frac{B}{a^{3(1+\alpha)}} \Big)^{\frac{1}{1+\alpha}}.
\end{equation}
Here $B$ is an integration constant. The corresponding state
parameter takes the form
\begin{equation}\label{38}
w=-A\Big(A+\frac{B}{a^{3(1+\alpha)}} \Big)^{-1},
\end{equation}
whose time derivative yields
\begin{equation}\label{39}
\dot w=-3ABH(1+\alpha)\Big(A+\frac{B}{a^{3(1+\alpha)}}\Big)^{-2},
\end{equation}
Using (\ref{38}) and (\ref{39}) in (\ref{18}) and (\ref{19}), we get
\begin{eqnarray}\label{40}
r&=&1+\frac{9}{2}AB\Omega_x(1-a^{-3(1+\alpha)})\Big(A+\frac{B}{a^{3(1+\alpha)}}\Big)^{-2}
-\frac{\Delta_G}{2}\Big[ 1-3A\Omega_x
\Big(A+\frac{B}{a^{3(1+\alpha)}}\Big)^{-1} \Big].
\end{eqnarray}
\begin{eqnarray}\label{41}
s&=&1-(A+B(1+\alpha))\Big(A+\frac{B}{a^{3(1+\alpha)}}\Big)^{-1}+\frac{\Delta_G}{9A\Omega_x
}\Big(A+\frac{B}{a^{3(1+\alpha)}} \Big)\Big[
1-3\Omega_xA\Big(A+\frac{B}{a^{3(1+\alpha)}} \Big)^{-1}
\Big].\nonumber\\
\end{eqnarray}

\section{Concluding remarks}

In this paper, we calculated the corrections to statefinder
parameters due to variable gravitational constant. These corrections
are relevant because several astronomical observations provide
constraints on the variability of $G$. An important thing to note is
that the $G-$corrected statefinder parameters are still geometrical
since the parameter $\Delta_G$ is a pure number and is independent
of the geometry.

\end{document}